\documentclass[12pt]{iopart}
\usepackage{iopams}
\usepackage{amssymb,graphicx,epsfig,iopams,multirow}

\begin{document}
\title[Importance of structural distortions in FeSe$_{1-x}$Te$_x$ superconductors]{Importance of structural distortions in enhancement of transition temperature in FeSe$_{1-x}$Te$_{x}$ superconductors}
\author{Kapil E. Ingle$^1$, K.R. Priolkar$^1$, Rayees A. Zargar$^2$, V. P. S. Awana$^3$ , S. Emura$^4$}
\address{$^1$ Department of Physics, Goa University, Goa 403206, India}
\address{$^2$ Department of Physics, Jamia Millia Islamia, New-Delhi, 110025, India}
\address{$^3$ National Physical Laboratory, Dr. K.S. Krishnan Marg, New Delhi 110012, India}
\address{$^4$ Institute of Scientific and Industrial Research, Osaka University, Osaka, Japan}
\ead{krp@unigoa.ac.in}

\begin{abstract}
Temperature (12K $\le$ T $\le$ 300K) dependent extended X-ray absorption fine structure (EXAFS) studies at the Fe K edge in FeSe$_{1-x}$Te$_x$ (x = 0, 0.5 and 1.0)  compounds have been carried out  to understand the reasons for increase in T$_C$ upon Te doping in FeSe. While local distortions are present near superconducting onset in FeSe and FeSe$_{0.5}$Te$_{0.5}$, they seem to be absent in non superconducting FeTe. Of crucial importance is the variation of anion height. In FeSe$_{0.5}$Te$_{0.5}$, near superconducting onset, the two heights, $h_{Fe-Se}$ and $h_{Fe-Te}$ show a nearly opposite behaviour. These changes indicate a possible correlation between Fe-chalcogen hybridization and the superconducting transition temperature in these Fe-chalcogenides. 
\end{abstract}

\pacs{ }
\vspace{2pc}

\submitto{\SUST}

\maketitle

\section{Introduction}
Superconductivity in F doped LaFeAsO (T$_C$ = 26K) \cite{kam} is considered ground breaking because it led to almost immediately, further discovery of many more Fe based superconductors, some with even higher T$_C$ \cite{SmAs}. Since this discovery in the so called 1111 type compounds, many other Fe containing pnictides and chalcogenides have been discovered to exhibit superconductivity. These include, BaFe$_{2}$As$_{2}$ (122 type) \cite{rott}, MFeAs (111) family \cite{wang}, FeSe (11) family \cite{hsu}, Sr$_{3}$MO$_{3}$FePn or the 21311 type \cite{ogino,zhu} and the defect structure A$_{0.8}$Fe$_{1.6}$Se$_{2}$ \cite{fe2se2}. These iron superconductors present themselves with glaring similarities in local structure around Fe and the electronic structure.  The main similarity in their structure is the quasi two dimensional Fe-Pn/Ch (Pn – Pnictogen, Ch – chalcogen atom) networks separated with or without a spacer layer.

Though the pairing mechanism in Fe based superconducting compounds is still not fully understood, it has been well established that properties of FePn/Ch superconductors are fundamentally different from conventional superconductors. The pairing mechanism could be related to coexisting magnetic interactions or spin fluctuations \cite{ins} or inter orbital pair hopping \cite{stewart}. Although there are many important differences between the FePn/Ch superconductors and cuprates, primarily in the symmetry of the superconducting gap, a close similarity in the phase diagram of these Fe superconductors and the cuprates and even the heavy fermion superconductors has been established \cite{scal}.

The quasi two dimension Fe-Pn/Ch networks consist of Fe-Pn/Ch tetrahedra whose angle as well as the height of the Pn/Ch ion above or below the plane of Fe ions play a crucial role in superconductivity of these compounds. The perpendicular distance between the plane of Fe atoms and the plane of Pn/Ch ions is defined as the Anion height ($h$). It has been found that $h$ has a great influence on T$_C$ as $h$ directly influences the hybridization between the Fe 3d bands and the Pn/Ch p bands \cite{sub}. Both hydrostatic and chemical pressures tend to influence $h$ and therefore the superconducting transition temperature \cite{rott,josep,granad}. This is further emphasized by pressure dependent Fe X-ray absorption fine structure (XAFS) studies on iron chalcogenide superconductors \cite{bend1,bend2}. These studies show even under hydrostatic pressure, T$_C$ is more influenced by local structural changes as is the case with chemical pressure \cite{josep2}.    

Recent XAFS studies at the As K edge as a function of temperature have shown a significant correlation between thermal variation of anion height and superconducting onset temperature in SmFeAsO$_{0.8}$F$_{0.2}$ \cite{kapil}. Such a correlation is absent in the non superconducting compound. Local structural parameters obtained from EXAFS measurements from the pnictogen/chalcogen site are generally more reliable. Furthermore, the increase in Fe-As bond distance just below the onset temperature has been related to local structural distortions observed around Cu ions in high T$_C$ cuprates \cite{oya}.   The correlation between local structural parameters and superconducting transition has also been highlighted using EXAFS in other isostructural systems \cite{zhang1,josep3}.

Among the iron based superconductors, FeSe, though exhibits the lowest superconducting transition temperature (T$_C$ $\sim$ 8K), is considered as a model system for understanding the superconductivity mechanism in these materials \cite{mizu}. Another interesting aspect is that the structure of FeSe consists of no spacer layers and requires no external doping to exhibit superconductivity. Furthermore, even though FeTe is non superconducting, substitution of Te for Se increases the T$_C$ to $\sim$ 15K. The phenomenon of disorder/impurity induced superconductivity is a unique property of these compounds. Additionally, superconductivity can also be induced by oxygen annealing \cite{sun} or by doping S in FeTe \cite{mizu2,si} . However, substitutions at the Fe site generally results in destruction of superconductivity \cite{mizu1,anuj}. This highlights the importance of hybridization between Fe $3d$ and Ch $p$ orbitals in superconducting properties of these compounds. Such $d-p$ hybridization critically depends on the local structure around Fe. Presence of disorder at a local structural level has been confirmed by X-ray absorption near edge structure (XANES) and EXAFS studies in FeSe$_{1-x}$Te$_{­x}$, T$_C$ \cite{josep,josep1,ida}. These studies indicate a phase separation characterized by different iron-chalcogen bond distances over nanoscopic length scale \cite{he}. However, the role of these nanoscopic phase separation in enhancing T$_C$ is still not clearly understood. It is also worth investigating the similarities if any, in local structural distortions near the onset of superconductivity between these 11 type Fe superconductors and other superconducting families like cuprates or 1111 type Fe based superconductors. To seek answers to the above questions and to establish a correlation between local structural distortions and superconductivity in FeSe type superconductors, we have carefully studied Fe K edge EXAFS as a function of temperature in superconducting FeSe and FeSe$_{0.5}$Te$_{0.5}$ and non-superconducting FeTe compounds.

\section{Experimental}
The bulk polycrystalline FeSe$_{1-x}$Te$_{x}$ samples with x = 0, 0.5 and 1.0 were synthesized through standard solid state reaction route via vacuum encapsulation. The high purity chemicals Fe, Se, and Te were weighed in the stoichiometric ratio and ground thoroughly in a Glove box having pure Argon atmosphere. The mixed powder was subsequently pelletized and then encapsulated in an evacuated (10$ ^{-3} $ Torr) quartz tube. The encapsulated tube was then heated at 750$^\circ$C for 12 hours and slowly cooled to room temperature. The heating schedule was repeated couple of times with intermediate grinding. The X-ray diffraction (XRD) pattern was recorded at room temperature in the scattering angular (2$\theta$) range of $ 10 ^\circ $ to $ 80^\circ $ in equal 2$\theta$ step of 0.02$^\circ$ using Rigaku Diffractometer with Cu K$_\alpha$ ($\lambda$ = 1.54$\AA$). Rietveld analysis was performed using the standard FullProf program. The resistivity measurements were performed on Quantum design Physical Property Measurements System (PPMS-14T, Quantum Design)

EXAFS measurements at the Fe K edge were performed in transmission mode at BL09C beamline at Photon Factory, Japan. EXAFS was scanned from -200 eV to 1000 eV with respect to Fe K edge energy (7112 eV).  Both incident (I$_0$) and transmitted (I) intensities were measured simultaneously using ionization chamber filled with appropriate gases. The absorbers were prepared by sprinkling finely ground powder on scotch tape and stacking several such layers to optimize the thickness so that the edge jump ($\Delta\mu$) was restricted to $\le$ 1.

EXAFS data analysis in the $k$ range of 2 to 14\AA$^{-1}$ and in the $R$ range of 1 to 3\AA~ was performed using Demeter program \cite{ravel}. Theoretical amplitude and phase shift functions for different correlations were calculated using FEFF 6.01 \cite{rehr} using the crystal structure data obtained from room temperature X-ray diffraction.

\section{Results and Discussion}
\begin{figure}
\centering
\includegraphics[width=\columnwidth]{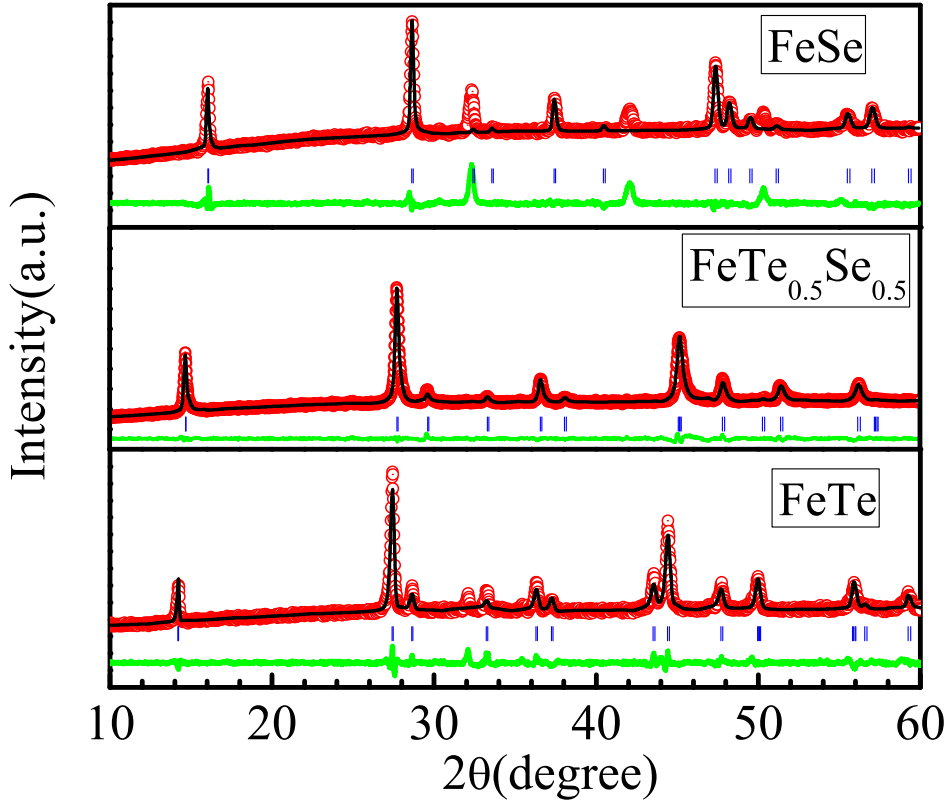}
\caption{\label{figure1} Room temperature XRD patterns of FeSe$_{1-x}$Te$_{x}$ (x = 0, 0.5 and 1)}
\end{figure}

Room temperature XRD patterns of FeSe$_{1-x}$Te$_{x}$ samples with x = 0, 0.5 and 1.0 are shown one over the other in \ref{figure1} . All the studied samples crystallize in tetragonal structure with space group P4/nmm. The lattice parameters $a$ and $c$ are respectively 3.82(2)\AA~ and 6.28(1)\AA~ for FeTe, 3.79(1)\AA~ and 6.02(2)\AA~ for FeTe$_{0.5}$Se$_{0.5}$ and 3.77(3)\AA~ and 5.57(1)\AA~ for FeSe. The increase in both $a$ and $c$ lattice parameters with increasing Te content is indicative of successful substitution of bigger ion Te at Se site. Furthermore, the obtained values of lattice parameters compare well with those reported in literature earlier \cite{hsu,mizu,braj}.  It may be mentioned that FeSe XRD contains hexagonal (NiAs type) P63/mmc (196) as a minority phase along with the majority tetragonal i.e., P4/nmm(129) \cite{braj}. 

\begin{figure}
\centering
\includegraphics[width=\columnwidth]{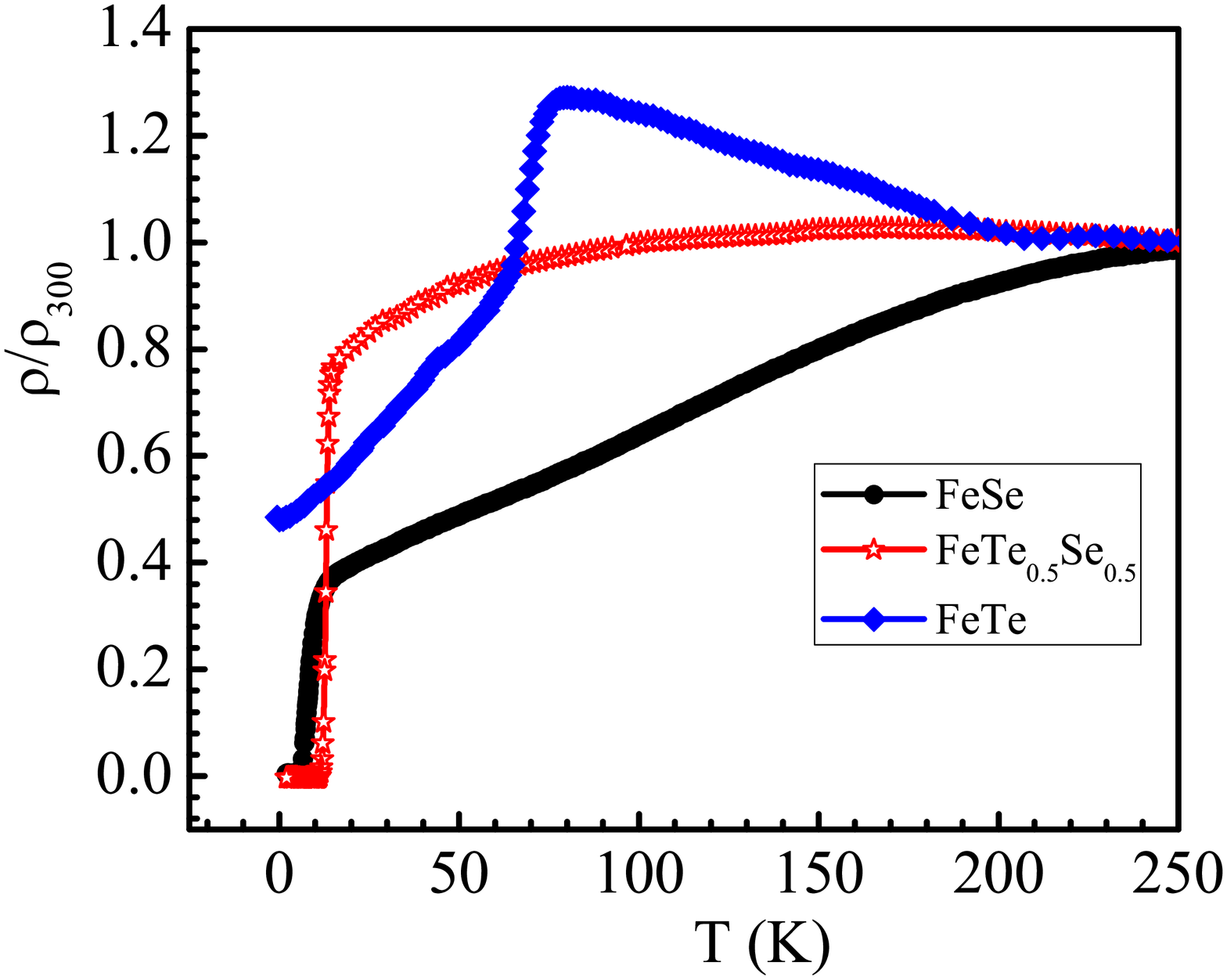}
\caption{\label{RES} Normalized resistivity as a function of temperature for FeSe $_{1-x}$ Te $_{x}$, $x$ = 0, 0.5 and 1.0}
\end{figure}

The resistivity versus temperature plots for studied FeSe $_{1-x}$Te$_{x}$ samples with x = 0, 0.5 and 1.0 are depicted in figure\ref{RES}. The FeTe compound exhibits a hump like metallic step in resistivity at around 80K, which is indicative of the magnetic ordering of Fe spins \cite{vps}. FeTe is not superconducting down to 2K. The FeSe compound is metallic in nature althrough from 300K down to 7K and becomes superconducting below this temperature. The exact T$_C$(R=0) is seen at 6K. The FeTe$_{0.5}$Se$_{0.5}$ is superconducting below 14K and is also metallic in nature in normal state. The resistivity value at room temperature is least for FeTe$_{0.5}$Se$_{0.5}$ and maximum for FeTe. Both the XRD and resistivity results are in general agreement with previous reports on these compounds viz. \cite{braj,vps,vps1,vps2}. 

\begin{figure}
\centering
\includegraphics[width=\columnwidth]{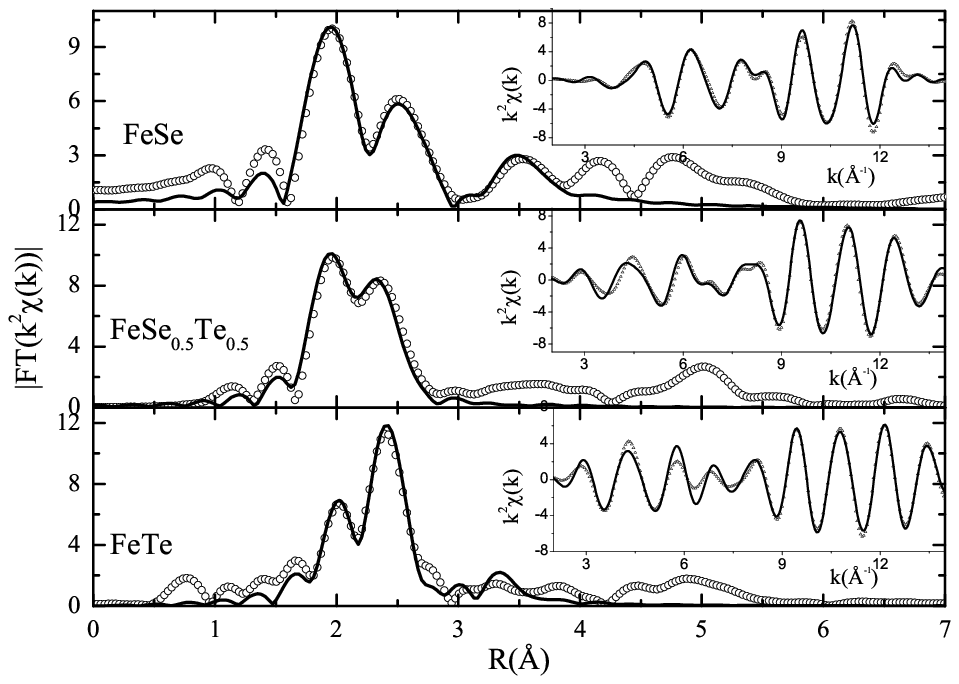}
\caption{\label{xafs} Magnitude of Fourier transform of $k^2$ weighted $\chi$(k) recorded at room temperature in FeSe, FeSe$_{0.5}$Te$_{0.5}$ and FeTe. The solid lines are best fits to the data. Inset shows Fourier filtered $k^2$ weighted XAFS data along with fitted curves for the three samples.}
\end{figure}

Magnitudes of the Fourier transform (FT) of EXAFS spectra recorded at 12K in the three compounds are presented in figure \ref{xafs} and compare well with earlier reports \cite{josep,bend1,ida}. The data for all three compounds exhibit a double peak structure in the range 1.5 to 3\AA. In case of FeSe, the first peak just below 2\AA~ corresponds to Fe-Se nearest neighbour correlation while the second peak around 2.5\AA~ arises due to Fe-Fe correlation. Changes in the FT spectra with changes in scattering atoms are clearly visible in the figure. Apart from the change in intensity ratio, the nearest neighbour correlation appears at a slightly higher distance in FeTe. The FT spectra of FeSe$_{0.5}$Te$_{0.5}$, on the other hand appears to be a linear addition of the two end members. The data in the range 1 to 3\AA~ was fitted to the first two correlations – Fe-Se/Te and Fe-Fe.  In the present fitting protocol, the bond distances of above correlations were expressed in terms of geometric relations based on lattice parameter ($a$) and anion height ($h$). In addition to $a$ and $h$, mean square relative displacement (MSRD or $\sigma^2$) were also varied for each path. Fe K edge EXAFS data along with fits at representative temperature (12K) in both R and k space is shown in Figure\ref{xafs}. The bond distances calculated from fitted values of $a$ and $h$ and corresponding $\sigma^2$ at few temperatures are listed in Table \ref{fese-tab1}.

\fulltable{\label{fese-tab1}Bond lengths and mean square relative displacements of near neighbours in FeSe, FeSe$_{0.5}$Te$_{0.5}$ and FeTe at selected temperatures.}
\br
\multirow{2}{*}{Temperature} & \multirow{2}{*}{Bond x C.N.} & \multicolumn{2}{c}{FeSe} & \multicolumn{2}{c}{FeSe$_{0.5}$Te$_{0.5}$} & \multicolumn{2}{c}{FeTe} \\ 
 &  & R (\AA) & $\sigma^2$\AA$^2$ & R (\AA) & $\sigma^2$\AA$^2$ & R (\AA) & $\sigma^2$\AA$^2$ \\ \mr
\multirow{3}{*}{300K} & Fe – Se $\times$ 4 & 2.39(1) & 0.006(1) & 2.36(1) & 0.005(1) & -- & --  \\  
 & Fe – Te $\times$ 4 & -- & --  & 2.58(1) & 0.004(1) & 2.63(1) & 0.006(1) \\  
 & Fe – Fe $\times$ 4 & 2.66(1) & 0.008(1) & 2.65(1) & 0.008(1) & 2.75(1) & 0.012(2) \\ \mr
\multirow{3}{*}{80K} & Fe – Se $\times$ 4 & 2.386(8) & 0.0048(10) & 2.388(8) & 0.0037(6) & -- & -- \\  
 & Fe – Te $\times$ 4 & -- & -- & 2.575(8) & 0.0024(6) & 2.582(6) & 0.0049(4) \\ 
 & Fe – Fe $\times$ 4 & 2.667(9) & 0.005(1) & 2.656(8) & 0.006(1) & 2.670(9) & 0.010(2) \\ \mr
\multirow{3}{*}{40K} & Fe – Se $\times$ 4 & 2.378(5) & 0.0043(10) & 2.388(4) & 0.0038(4) & -- & -- \\  
 & Fe – Te $\times$ 4 & -- & -- & 2.568(4) & 0.0023(6) & 2.581(6) & 0.0043(3) \\  
 & Fe – Fe $\times$ 4 & 2.656(8) & 0.005(1) & 2.660(8) & 0.005(1) & 2.672(9) & 0.009(1) \\ \mr
\multirow{3}{*}{20K} & Fe – Se $\times$ 4 & 2.378(5) & 0.0039(8) & 2.387(4) & 0.0029(4) & -- & -- \\  
 & Fe – Te $\times$ 4 & -- & -- & 2.573(4) & 0.0039(10) & 2.577(6) & 0.0042(3) \\ 
 & Fe – Fe $\times$ 4 & 2.654(7) & 0.005(1) & 2.663(7) & 0.006(1) & 2.668(9) & 0.009(1) \\ \mr
\multirow{3}{*}{12K} & Fe – Se $\times$ 4 & 2.383(5) & 0.0039(6) & 2.386(4) & 0.0031(7) & -- & -- \\  
 & Fe – Te $\times$ 4 & -- & -- & 2.559(4) & 0.0045(10) & 2.577(6) & 0.0045(3) \\ 
 & Fe – Fe $\times$ 4 & 2.661(5) & 0.005(1) & 2.654(5) & 0.006(2) & 2.668(9) & 0.009(1) \\ \br
\endfulltable


FeSe and FeTe are known to undergo a phase transition from the room temperature tetragonal phase to a lower symmetry phase \cite{mar,pom,marti}. As a result the Fe-Fe bond distance at 2.66\AA~ which has a degeneracy of 4 splits into two doubly degenerate correlations at 2.57\AA~ and 2.7\AA. Such a splitting in Fe-Fe bond distance is also reported in previous EXAFS analysis \cite{josep}. However, in the present case the near neighbour distances are extracted from $a$ and $h$ and hence only one, four fold degenerate Fe-Fe correlation along with Fe-Se/Te correlation have been used to fit the low temperature EXAFS data. The variation of Fe-Fe bond distances as a function of temperature in the range 10K $\leq$ T $\leq$ 80K along with Fe-Se and Fe-Te bond distances in all three compounds is presented in figure \ref{xafs}. A narrow variation of Fe-Fe bond length  can be seen in all three compounds.

\begin{figure}
\centering
\includegraphics[width=\columnwidth]{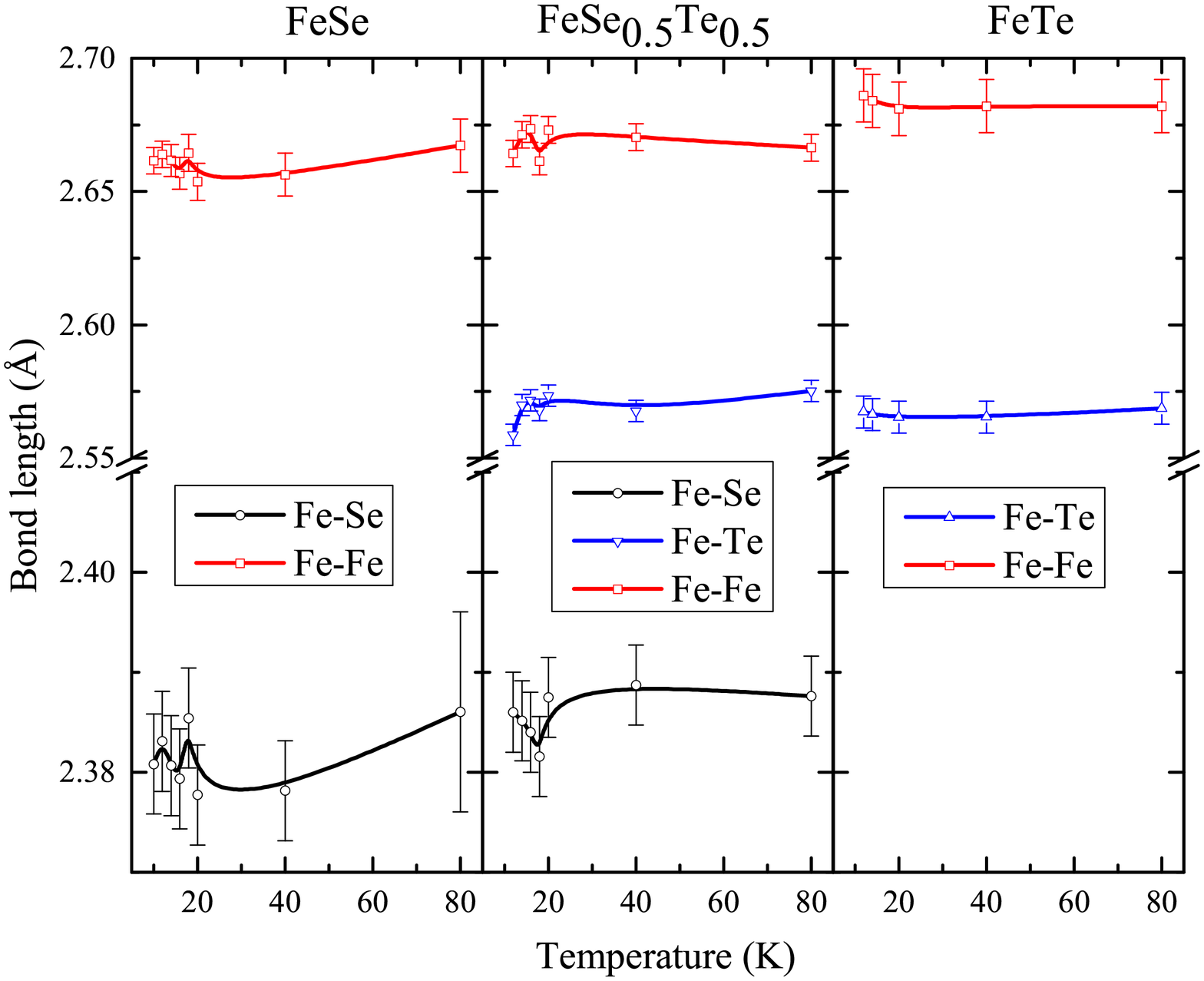}
\caption{Temperature variation of Fe-Se, Fe-Te and Fe-Fe bond distances obtained from EXAFS analysis in FeSe,  FeSe$_{0.5}$Te$_{0.5}$ and FeTe.}
\end{figure}

The Fe-Se/Te bond distances also exhibit a very narrow variation. However, one can notice two distinct features in the superconducting compounds, FeSe and FeSe$_{0.5}$Te$_{0.5}$ as opposed to FeTe even though they are within error bars. Firstly, the Fe-Se bond length in FeSe shows an anomalous behaviour below 18K. Similarly a weak rise after a distinct minimum is noticed in FeSe$_{0.5}$Te$_{0.5}$ just below 20K which agrees well with its T$_C$-onset = 18K. Infact, a small minimum can also be seen at about 14K in FeSe. Though these features are buried within statistical error bars, presence of local structural distortions cannot be eliminated. Recently, such anomalies in nearest neighbour Fe-As bond distances have been shown to be present F doped SmFeAsO superconductors \cite{kapil}. Distortions in Cu-O  bond distances has also been observed near the onset of superconductivity in cuprate superconductors. It must also be mentioned here that Fe-Se and Fe-Te bond lengths in FeSe$_{0.5}$Te$_{0.5}$ are close to their values in respective undoped compounds, FeSe and FeTe and not equal to the one calculated from average structure obtained from Rietveld refinement of XRD data.  This observation is consistent with previously reported EXAFS analysis \cite{josep,ida}. Secondly, the temperature variations of Fe-Se bond distance and Fe-Te bond distance are exactly opposite to each other indicating critical changes in Fe $3d$­ Ch $p$ hybridization. Recent photoemission studies have indeed shown similar changes in contribution of chalcogen ion to the density of states near Fermi level in FeSe0.6Te0.4 \cite{kalo}. In contrast, the Fe-Te bond length shows very little or no temperature variation below 20K. These observations lend weight to the possibility of presence of local structural distortions in these Fe-chalcogenide superconductors near their T$_{C-onset}$.

\begin{figure}
\centering
\includegraphics[width=\columnwidth]{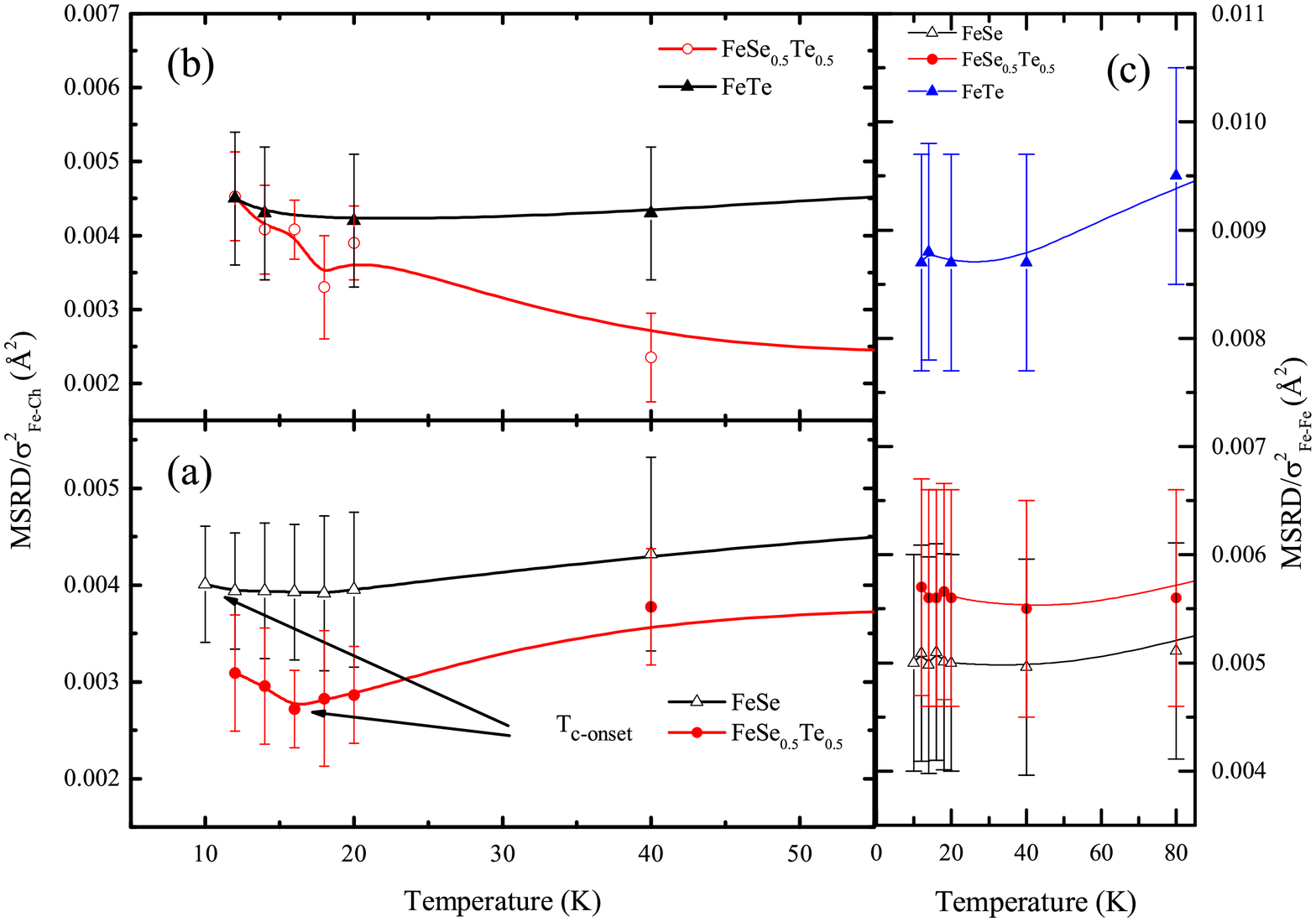}
\caption{\label{msrd} Mean square relative displacement in (a) Fe-Se and (b) Fe-Te and (c) Fe-Fe bond distances as a function of temperature in FeSe, FeSe$_{0.5}$Te$0.5$ and FeTe.}
\end{figure}

In order to confirm the variations seen in Fe-Se and Fe-Te bond distances in superconducting compound are due to structural distortion and not merely an artifact of nonlinear fitting, variation of $\sigma^2$ of Fe-Se and Fe-Te bonds in all the compounds is presented in Fig. \ref{msrd} It can be seen from Figure \ref{msrd} that MSRD of Fe-Se bond in  FeSe$_{0.5}$Te$_{0.5}$ is larger in magnitude than that in FeSe. Similar behaviour is noticed in case of Fe-Te bond in  FeSe$_{0.5}$Te$_{0.5}$ and FeTe.  The structural disorder induced due to partial replacement of Se by Te or vice versa is responsible for such behaviour. Furthermore, the $\sigma^2$ values corresponding to Fe-Se bonds exhibit an upturn peaking at around 16K in  FeSe$_{0.5}$Te$_{0.5}$ and at about 14K in FeSe. These temperatures correspond well with the respective T$_C$-onset of the two compounds. Similar behaviour was also seen in fluorinated RFeAsO \cite{kapil,zhang}. In these arsenide superconductors the upturn temperature was identified with characteristic temperature T* or the pseudogap temperature observed in cuprates.  In the present case the upturn temperature for both the compounds is found to be about 20K. Although there are not enough measurements in the temperature range 20K to 40K to pinpoint this temperature, the ratio of upturn temperature to peak temperature as deduced from present measurements is about 1.4 to 1.6 which is in good agreement with that obtained for arsenide superconductors \cite{kapil}. This observation further cements the similarities in superconducting mechanism between iron based superconductors and high T$_C$ cuprates. It may be mentioned here that Fe-Se bond distances and MSRD values could have been more reliably obtained especially in FeSe$_{0.5}$Te$_{0.5}$ from a temperature dependent study at the Se K edge. 

The deviation of $\sigma^2$  of Fe-Se bond from the expected Debye behavior in superconducting FeSe and FeSe$_{0.5}$Te$_{0.5­}$ confirms the anomaly obtained in Fe-Se bond distances of the two compounds to be due to some kind of  dynamical structural distortions as the temperature is decreased below T* to approach the superconducting transition. 

\begin{figure}[h]
\centering
\includegraphics[width=\columnwidth]{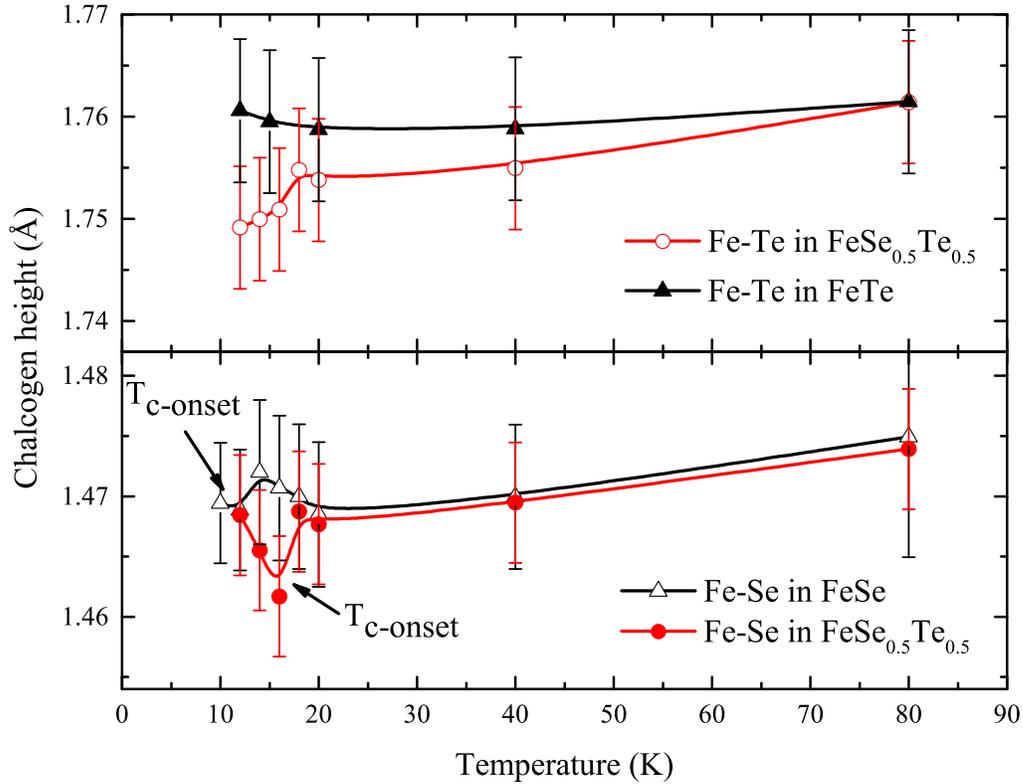}
\caption{\label{height} Chalcogen height ($h$) as a function of temperature in FeSe, FeSe$_{0.5}$Te$0.5$ and FeTe.}
\end{figure}

The structural distortions evident from trends of Fe-Se/Te bond length and $\sigma^2$ as a function of temperature though point towards an important similarity between these 11 type Fe based superconductors and other families of superconducting compounds, but still leaves the question of enhancement of T$_C$ on Te doping in FeSe unanswered. In order to explore the reason behind this increase in T$_C$, in Figure\ref{height}, thermal evolution of anion height $h$ is plotted in all three compounds. Anion height is considered to be an important parameter for superconductivity in Fe based compounds as it directly related to the hybridization between Fe $3d$ and chlacogen $p$ bands \cite{anuj,chen}. Greater the hybridization or smaller the values of $h$, higher is the mobility of holes, thereby suppressing the antiferromagnetic interactions and ushering in superconductivity. This fact is quite clearly noticed in values of $h$ obtained from the fittings to the present EXAFS data as well as by earlier workers \cite{josep,ida}. While the value of $h$ in FeTe was around 1.75\AA, that in superconducting FeSe was found to be 1.47\AA . Though this value of $h$ is much smaller than that in non-superconducting compounds, it is larger compared to that obtained in case of fluorinated SmFeAsO \cite{kapil}. The other difference that can be directly noted is the difference in values of $h$ in superconducting and non superconducting compounds of 11 type chalcogenides and those belonging to 1111 type arsenides. While in case of arsenides the difference was found to be about 0.006\AA, in case of 11 type chalcogenides it is nearly 50 times ($\sim$ 0.28\AA).  This could be related to the nature of substitution. In 1111 type compounds the substitution is in REO layers (RE – rare earth) and therefore only indirectly affects the Fe-As bonds while here Se is replaced by Te. As has been previously reported, in case of FeSe$_{0.5}$Te$_{0.5}$, $h$ for Fe-Se bond is similar to that in FeSe and likewise for Fe-Te bond. This implies that both Se and Te tend to preserve their structural environment found in FeSe and FeTe even in the doped compound. A similar observation has been also made using scanning, tunneling microscopy \cite{he}. 

The temperature variation of $h$ reveals interesting trends. Although the value of $h$ for Fe-Se bond in FeSe and FeSe$_{0.5}$Te$_{0.5}$ are very similar at room temperature (Figure \ref{height}), at lower temperatures, especially below 20K, the values of $h$ in  FeSe$_{0.5}$Te$_{0.5}$ are slightly lower. Further $h$ has a local minimum near the superconducting onset temperatures of both  FeSe$_{0.5}$Te$_{0.5}$ and FeSe. Similar minimum was also obtained in SmFeAsO$_{1-x}$F$_x$ superconductors \cite{kapil}.  Such a minimum in $h$ around T$_{C-onset}$ points towards the presence of an electronic instability around this temperature. A similar behaviour of Cu-O peak width was also observed in high T$_C$ cuprates \cite{mustre}.  It is also interesting to note that anion height of Fe-Te-Fe network ($h_{Fe-Te}$) in  FeSe$_{0.5}$Te$_{0.5}$ presents an exactly complementary behaviour. At low temperatures, when $h_{Fe-Se}$ (anion height of Fe-Se-Fe network), exhibits a local minimum, $h_{Fe-Te}$ presents a maximum. Further, while $h_{Fe-Se}$ increases, $h_{Fe-Te}$ shows a decrease. No such temperature variation is seen in the temperature variation of anion height of FeTe. This is a very important result as the lowering of $h_{Fe-Se}$ at the expense of $h_{Fe-Te}$ would mean an increase in hybridization between the $3d$ orbitals of Fe and $4p$ orbitals of Se concomitant with a decrease in hybridization between Fe $3d$ and Te $5p$. This fact along with slightly lower values of $h_{Fe-Se}$ in FeSe$_{0.5}$Te$_{0.5}$ as compared to FeSe is perhaps responsible for a higher T$_C$ in  FeSe$_{0.5}$Te$_{0.5}$.

Present results not only point towards similarities in the mechanism of superconductivity in other families of Fe based superconductors but also hint towards an intimate connection in the mechanism of superconductivity in Fe based superconductors and that proposed in cuprates. This along with the similarities in their phase diagrams, the proximity of antiferromagnetic order and superconductivity and the presence of spin resonance peak in superconducting region indicates unconventional nature of superconductivity. Furthermore, the complementary behaviour of $h_{Fe-Se}$ and $h_{Fe-Te}$ near the superconducting onset temperature in FeSe$_{0.5}$Te$_{0.5}$ indicates that a strong hybridization between Fe $3d$ and Se $4p$ bands is a necessary condition to free the localized holes to induce superconducting ground state. This strong $d-p$ hybridization could also be the reason for the increase in T$_C$ in spite of impurity doping.  It must be mentioned that variation of anion height indicates the doped impurity infact helps in strengthening the hybridization between Fe 3d and Se 4p bands.  Therefore the anion height becomes the important parameter in controlling the T$_C$. Indeed, the present investigations also show that for T $\le$ 20K, the $h_{Fe-Se}$ in FeSe$_{0.5}$Te$_{0.5}$ is slightly smaller than that in pure FeSe agreeing well with the observed increase in T$_C$ with Te doping.

\section{Conclusions}
Temperature dependent EXAFS studies carried out at the Fe K edge in FeSe$_{1-x}$Te$_x$ (x = 0, 0.5 and 1.0)  compounds indicate that significant local distortions are present in FeSe and FeSe$_{0.5}$Te$_{0.5}$ near the superconducting onset. Such distortions seem to be absent in non superconducting FeTe. Particularly interesting is the variation of anion height. In FeSe$_{0.5}$Te$_{0.5}$, the phase separation at the local level results in two anion heights, $h_{Fe-Se}$ and $h_{Fe-Te}$. Near superconducting onset, the two heights show nearly opposite behaviour. Such behaviour could be perhaps due to a correlation between Fe-chalcogen hybridization and superconductivity in Fe chalcogenides.

\ack{Authors thank Department of Science and Technology, Govt. of India for financial assistance under the project SR/S2/CMP-57. Thanks are also due to Photon Factory for beamtime under the proposal 2011G0077.}


\Bibliography{100}
\bibitem{kam}Kamihara Y, Watanabe T, Hirano M and Hosono H 2008 J. Am. Chem. Soc. \textbf{130} 3296
\bibitem{SmAs} Wu G et al 2009 J. Phys.: Condens. Matter \textbf{21} 142203
\bibitem{rott}Rotter M, Tegel M and Johrendt D 2008 Phys. Rev. Lett. \textbf{101} 107006
\bibitem{wang}Wang C, Li Y K, Zhu Z W, Jiang S, Lin X, Luo Y K, Chi S, Li L J, Ren Z, He M, Chen H, Wang Y T, Tao Q, Cao G H and Xu Z A 2009 Phys. Rev. B \textbf{79} 054521
 \bibitem{hsu}Hsu, F. C. et al. 2008 Proc. Natl. Acad. Sci. USA \textbf{105}, 14262
\bibitem{ogino}H. Ogino et al 2009 Supercond. Sci. Technol. \textbf{22} 085001
\bibitem{zhu}Zhu X, Han F, Mu G, Cheng P, Shen B, Fang L, and Wen H H 2009 Phys. Rev. B \textbf{79} 220512
\bibitem{fe2se2} Bao, W, Li G N, Huang Q, Chen G F, He J B, Green M A, Qiu M, Wang D M, Luo and Wu  M M 2013 Chin. Phys. Lett. \textbf{30} 027402
\bibitem{ins} Onari, S, Kontani H, and Sato M, 2010 Phys. Rev. B \textbf{81} 060504
\bibitem{stewart} Stewart G. R. 2011 Rev. Mod. Phys. \textbf{83} 1589
\bibitem{scal} Scalapino D. J. 2012, Rev. Mod. Phys. \textbf{84} 1383
\bibitem{sub}Subedi A, Zhang L, Singh D, Du M 2008 Phys. Rev. B \textbf{78} 134514
\bibitem{josep}Joseph B, Iadecola A, Puri A, Simonelli L, Mizuguchi Y, Takano Y and Saini N L 2010 Phys. Rev. B \textbf{82} 020502
\bibitem{granad}Granado E {\it et al} 2011 Phys. Rev. B \textbf{83} 184508 
\bibitem{bend1}Bendele M, Marini C, Joseph B, Simonelli L, Dore P, Pascarelli S, Chikovani M, Pomjakushina E, Conder K, Saini N L and Postorino P (2013) J. Phys.: Condens. Matter \textbf{25} 425704
\bibitem{bend2} Bendele M {\it et al} 2013 Phys. Rev. B {\textbf 88} 180506 
\bibitem{josep2} Joseph B {\it et al} 2013 Supercond. Sci. Technol. {\textbf 26} 065005
\bibitem{kapil}Ingle K, Priolkar K R, Pal A, Awana V P S and Emura S 2014 Supercond. Sci. Technol. \textbf{27} 075010
\bibitem{oya}Oyanagi H, Zhang C, Tsukada A and Naito M 2008 J. Phys.: Conf. Series \textbf{108} 012038
\bibitem{zhang1}Zhang C J, Oyanagi H, Sun Z H, Kamihara Y and Hosono H 2010 Phys. Rev. B \textbf{81} 094516
\bibitem{josep3}Joseph B, Iadecola A, Malavasi L and Saini N L 2011 J. Phys. Condens. Matter \textbf{23} 265701
\bibitem{mizu} Mizuguchi, Y., and Y. Takano, 2010, J. Phys. Soc. Jpn. \textbf{79}, 102001
\bibitem{sun} Sun Y {\it et al}, 2014, Nature Scientific Rep. \textbf{4}, 4585
\bibitem{mizu2} Mizuguchi Y, Deguchi K, Kawasaki Y, Ozaki T, Nagao M, Tsuda S, Yamaguchi T and Tanako Y 2011 J. Appl. Phys. \textbf{109}, 013914
\bibitem{si} Si W, Jie Q, Wu L, Zhou J, Gu G, Johnson P D, and Li Q 2010 Phys. Rev. B \textbf{81} 092506
\bibitem{mizu1} Mizuguchi Y, Tomioka F, Tsuda S, Yamaguchi T and Takano Y 2009 J. Phys. Soc. Jpn. \textbf{78} 074712
\bibitem{anuj} Kumar A, Tandon R P and Awana V P S 2012 IEEE Trans. Magnetics 2012 
\bibitem{josep1} Joseph B, Iadecola A, Simonelli A, Mizuguchi Y, Takano Y, Mizokawa T and Saini N L 2010 J. Phys. Condens. Matter \textbf{22} 485702
\bibitem{ida} Idecola A Joseph B, Puri A, Simonelli L, Mizuguchi Y, Testemale D,  Proux O, Hazemann J-L, Takano Y and Saini N L 2011 J. Phys. Condens. Matter \textbf{23} 425701
\bibitem{he} He X, Li G, Zhang J, Karki A B, Jin R, Sales B C, Sefat A S, McGuire M A, Mandrus D and Plummer E W 2011 Phys. Rev. B \textbf{83} 220502
\bibitem{ravel} Ravel B and Newville M, 2005 J. Synchr. Rad. \textbf{12} 537
\bibitem{rehr} Zabinsky S I, Rehr J J, Ankudinov A, Albers A C and Eller M J 1995 Phys. Rev. B \textbf{52} 2995
\bibitem{braj}Tiwari B, Jha R, Awana V PS 2011 AIP Adv. \textbf{4} 067139
\bibitem{vps}Awana V P S, Pal A, Vajpayee A, Gahtori B, Kishan H 2011 Physica C \textbf{471} 77–82
\bibitem{vps1}Awana V P S, Pal A, Vajpayee A, Mudgel M, Kishan H, Husain M, Zeng R, Yu S, Guo Y F, Shi Y G, Yamaura K and Takayama-Muromachi E 2010 J. Appl. Phys. \textbf{107} 09E128
\bibitem{vps2}Awana V P S, Govind, Pal A, Gahtori B, Kaushik S D, Vajpayee A, Kumar J and Kishan H 2011 J. Appl. Phys. \textbf{109} 07E122
\bibitem{pom} Pomjakushina E, Conder K, Pomjakushin V, Bendele M and Khasanov R 2009 Phys. Rev. B \textbf{80} 024517
\bibitem{mar} Margodonna S, Takabayashi Y, McDonald M T, Karperkeiwics K, Mizuguchi Y, Takano Y, Fitch A N, Suard E and Prassides K 2008 Chem. Commun. 5607
\bibitem{marti}A. Martinelli, A. Palenzona, M. Tropeano, M. Putti, M. R. Cimberle, T. D. Nguyen, M. Affronte, and C. Ritter 2010 Phys. Rev. B \textbf{81} 094115
\bibitem{kalo} Adhikary G, Biswas D, Sahadev N, Ram S, Kanchana V, Yadav C S, Paulose P L and Maiti K 2013 J. Appl. Phys. \textbf{114} 163906
\bibitem{chen}Chen C L, {\it et al.} 2011 Phys. Chem. Chem. Phys. \textbf{13} 15666
\bibitem{zhang}Zhang C J, Oyanagi H, Sun Z H, Kamihara Y and Hosono H 2008 Phys. Rev. B \textbf{78} 214513
\bibitem{mustre} Mustre de Leon J, Conradson S D, Batistic I and Bishop A R 1990 Phys. Rev. Lett. \textbf{65} 1675
\endbib
\end{document}